\documentclass[prl,twocolumn,amsmath,amssymb,aps,superscriptaddress]{revtex4}
\usepackage{amssymb}
\usepackage{amsmath}
\usepackage{latexsym}
\usepackage{graphicx}

\def\argmax{\mathop{\rm argmax}}

\def\choose#1#2{\genfrac{(}{)}{0pt}{}{#1}{#2}}
\begin{document}
\title{Optimal ratio between phase basis and bit basis in QKD}
\author{Masahito Hayashi}
\email{hayashi@math.is.tohoku.ac.jp}
\address{Graduate School of Information Sciences, Tohoku University, Aoba-ku, Sendai, 980-8579, Japan}
%\pacs{03.65.Wj,03.65.Ud,02.20.-a}
%02.20.-a 	Group theory 
%(for algebraic methods in quantum mechanics, see 03.65.Fd; 
%for symmetries in elementary particle physics, see 11.30.-j)
%03.65.Ta Foundations of quantum mechanics; measurement theory 
%(for optical tests of quantum theory, see 42.50.Xa)
%03.65.Ud Entanglement and quantum nonlocality 
%(e.g. EPR paradox, Bell's inequalities, GHZ states, etc.) 
%(for entanglement production in quantum information, see 03.67.Mn; 
%for entanglement in Bose-Einstein condensates, see 03.75.Gg)
%03.65.Wj State reconstruction, quantum tomography
%03.67.-a 	Quantum information
%03.67.Dd 	Quantum cryptography
%03.67.Hk 	Quantum communication
%05.30.Jp 	Boson systems 

\begin{abstract}
In the original BB84 protocol, the bit basis and the phase basis are used with equal probability.
Lo et al (J. of Cryptology, 18, 133-165 (2005)) proposed to modify the ratio between the two bases by increasing the final key generation rate.
However, the optimum ratio has not been derived.
In this letter, in order to examine this problem, the ratio between the two bases is optimized for exponential constraints given Eve's information distinguishability and the final error probability.
\end{abstract}
\maketitle
%\section{Introduction}
Bennett \& Brasserd \cite{BB84} proposed the BB84 protocol for quantum key distribution.
It was shown that this protocol generates secret random bits between two distinct parties
even though the quantum channel has noise\cite{mayer1,SP}.
Once a security proof was obtained for this protocol, many researchers \cite{RGK,SRS,Chau,WMYK,Renner,GL,wang2,wang3,WMU2,BA} improved the key generation rate.
Lo et al \cite{Lo} proposed to improve the key generation rate by modifying the ratio between
the bit($+$) basis and the phase($\times$) basis.
In the original BB84 protocol, the sender Alice and the receiver Bob choose the $+$ basis and the $\times$ basis with equal probability.
However, this equal ratio is not essential, because the purpose of a random basis choice is
estimating the phase error rate in the channel of the qubits in the coincidence basis.
That is, in order to generate the secure keys from the raw keys with the $+$ basis,
it is sufficient to estimate the $+$ error rate precisely.
The aim of the present letter is to improve the key generation rate by modifying the ratio between two bases.

For example, the following protocol improves the key generation rate. 
When Alice and Bob communicate $N$ qubits, Alice and Bob use the $\times$ basis only in the randomly chosen $\sqrt{N}$ qubits and use the $+$ basis in the remaining $N-\sqrt{N}$ qubits.
In the above protocol, when the length of the code $N$ is sufficiently large, 
Alice and Bob can estimate the phase error rate precisely. 
Since $\sqrt{N}/N$ approaches zero, the rate of discarded qubits approaches zero.
That is, it is possible that the generation rate of the raw keys with transmitted qubits is almost 100\%.
Hence, in order to optimize this ratio, we have to choose a suitable formulation.
Due to the difficulty of the formulation, this optimization has not been treated in existing researches.
As a possible formulation, one may consider the optimization of the final key generation rate with a constant constraint for Eve's information in the finite length code.
However, as is discussed by 
Lo et al\cite{Lo}, Hayashi \cite{H}, and Scarani and Renner \cite{SR}, the formula of finite length code is not simple.
Furthermore, its analysis depends on the length of the code.

This letter focuses on the exponential constraint as an intermediate criterion between the finite-length case and the infinite-length case.
Exponential rate is a common measure in information theory\cite{Gal},
and it was discussed in QKD by several papers \cite{Ham,WMU,H}.
That is, we treat exponential constraints for the block error probability for final keys and 
for Eve's information distinguishability for final keys\cite{BHLMO}.
This letter optimizes the final key generation rate based on the key distillation protocol given by Hayashi \cite{H,H2}.
In this key distillation protocol, first, a classical error correction is performed.
Next, privacy amplification using the Toeplitz matrix, which is an economical random matrix\cite{Carter,Krawczyk}, is performed.
Hence, Eve's information distinguishability can be characterized by the phase error probability of the corresponding CSS code.

For a simpler analysis, a single photon source and the lossless quantum channel are assumed to be available.
Furthermore, as an ideal assumption, 
the random coding and the maximum likelihood decoding are assumed to be performed in the classical error correction part.
%Practically, LDPC code is used for an actual code in the classical error correction part\cite{HHHT2}, but its error probability approaches zero only with polynomial speed\cite{MB}.
%Thus, an ideal classical error correction with the random coding is assumed.

\vspace{2ex}
%\section{Result}
\noindent
[Result]\par
This letter focuses on the asymmetric protocol, in which,
Alice and Bob use the $\times$ basis with ratio $p_2$,
and they announce the check bits, which are randomly chosen with ratio $p_1$ among  
the bits whose Alice's basis and Bob's basis are the $+$ basis.
(As shown later, the optimal case is given when
the ratio of $\times$ basis by Alice is equal to that used by Bob.)
The performance of the protocol is characterized by 
the final error probability of the classical error correction and Eve's information distinguishability.
The latter is equal to $\|\rho_{AE}-\rho_{A}\otimes \rho_{E}\|_1$ for 
Eve's final state $\rho_{E}$, 
Alice's final state $\rho_{A}$, 
and the final state $\rho_{AE}$ of the joint system of the final keys.

Both quantities depend on the number $N$ of transmitted qubits, the observed error rates $q_+$ of $+$ basis, and the observed error rates $q_\times$ of $\times$ basis.
However, both quantities cannot be determined by the above values because these depend on Eve's attack.
Hence, it is possible to compute only both upper bounds, i.e., the upper bound $B_b(N,p_1,p_2,q_+)$ of the final error probability and the upper bound $B_p(N,p_1,p_2,q_{\times})$ of Eve's information distinguishability\cite{H,H2}, which do not depend on Eve's attack.
For a given constant $C$,
the following exponential constraint is considered:
\begin{align}
\lim_{N\to \infty}\frac{-1}{N}\log B_b(N,p_1,p_2,q_+) 
&\ge C
\label{e1}\\
\lim_{N\to \infty}\frac{-1}{N}\log B_p(N,p_1,p_2,q_{\times}) 
&\ge C.
\label{e2}
\end{align}
Then, the main target is the calculation of the rates 
$p_1$ and $p_2$ optimizing the final key generation rate
$R_{A}(p_1,p_2,q_+,q_{\times},C)$
under the conditions (\ref{e1}) and (\ref{e2}),
when $q:=q_+=q_{\times}$.

In this letter, these values are numerically calculated using the logarithm base $2$ with $C=0.0001$.
For example, when $N=100,000$, $2^{-C N}=2^{-10}$.
However, since the quantity $B_p(N,p_1,p_2,q_{\times})$ has some polynomial factor, it is greater than $2^{-10}$.

\begin{figure}[htbp]
\begin{center}
\includegraphics[width=8cm]{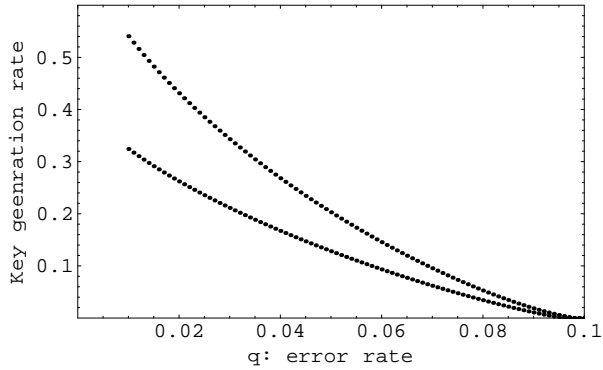}
\end{center}
\caption{Asymptotic key generation rate:
The upper line shows $\max R_{A}$, while
the lower line shows $\max R_{S}$.
}
\label{rate}
\end{figure}

Next, we consider the symmetric protocol, in which $\times$ basis is chosen with the same ratio with the $+$ basis in both sides.
In this case, it is possible to control only the ratio $p_1$ of the check bits.
The original BB84 protocol is given in this case as $p_1=1/2$.
%we call this case the modified BB84 protocol.
Then, we numerically calculate the rate $p_1$ optimizing the final key generation rate $R_{S}(p_1,q_+,q_{\times},C)$ under the conditions (\ref{e1}) and (\ref{e2}), when $q:=q_+=q_{\times}$.
The numerical results of $\max_{0\le p_1,p_2\le 1/2}R_{A}(p_1,p_2,q,q,0.0001)$ and 
$\max_{0\le p_1\le 1/2}R_{S}(p_1,q,q,0.0001)$ are shown in Figure \ref{rate}.

\begin{figure}[htbp]
\begin{center}
\includegraphics[width=8cm]{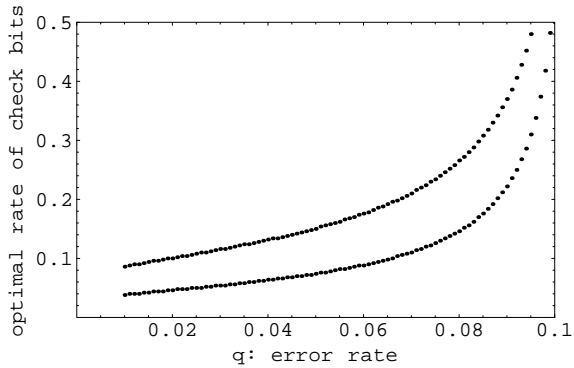}
\end{center}
\caption{Optimal choice rate for the check bits:
The upper line is $\argmax R_{S}$, while
the lower line is $\argmax R_{A}$.
}
\label{argmax1}
\end{figure}

Using the above results, we obtain Fig. \ref{argmax1}, which shows
$\argmax_{0\le p_1\le 1/2} \max_{0\le p_2\le 1/2} R_{A}(p_1,p_2,q,q,0.0001)$ and
$\argmax_{0\le p_1\le 1/2} R_{S}(p_1,q,q,0.0001)$. 

\begin{figure}[htbp]
\begin{center}
\includegraphics[width=8cm]{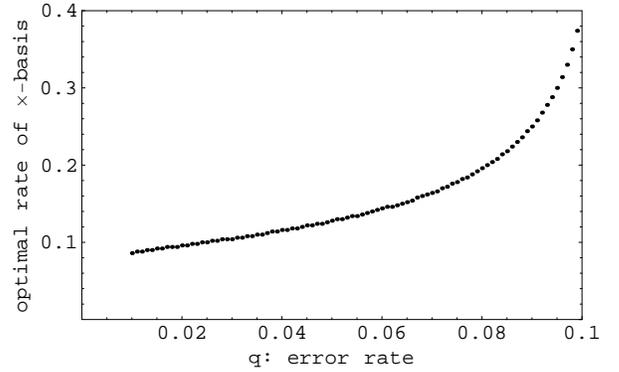}
\end{center}
\caption{Optimal choice rate of $\times$ basis}
\label{argmax2}
\end{figure}

Fig. \ref{argmax2} shows  
$\argmax_{0\le p_2\le 1/2} \max_{0\le p_1\le 1/2} R_{A}(p_1,p_2,q,q,0.0001)$.

\vspace{2ex}
\noindent
[Derivation]\par
%\section{Derivation}
The above figures are derived by combining the type method \cite{CK} and the analysis in Hayashi\cite{H}
as follows.
In this letter, we discuss the security based on the key distillation protocol given in Hayashi\cite{H},
in which, after the generation of raw keys, the classical error correction is performed using a pseudo-classical noisy channel, and random privacy amplification is applied using the Toeplitz matrix.
Similar to Hayashi \cite{H}, the following derivation focuses on the discrete (partial)-twirled channel, by considering the phase error rate and the bit error rate of this channel corresponding to the raw keys.

Let us calculate the probability that
the estimated phase error rate is $q_{\times}$, and the phase error rate among raw keys is $q_{\times}'$.
As discussed in Hayashi\cite{H},
this probability can be evaluated using the hypergeometric distribution, that is,
$(\choose{N p_2^2}{N p_2^2 q_\times }\choose{N (1-p_2)^2(1-p_1)}{N (1-p_2)^2(1-p_1)}q_{\times}')
/\choose{N (p_2^2+(1-p_2)^2(1-p_1))}{N p_2^2 q_\times+ N (1-p_2)^2(1-p_1)q_{\times}'}
$,
where $N$ is the total number of transmitted qubits.
Since $\frac{1}{n+1}2^{n h(k/n)} \le \choose{n}{k}\le 2^{n h(k/n)}$,
this probability is bounded by
$
(\choose{N p_2^2}{N p_2^2 q_\times }\choose{N (1-p_2)^2(1-p_1)}{N (1-p_2)^2(1-p_1)}q_{\times}')
/\choose{N (p_2^2+(1-p_2)^2(1-p_1))}{N p_2^2 q_\times+ N (1-p_2)^2(1-p_1)q_{\times}'}
\le 
\epsilon_p(q_{\times}'):=
\frac{
2^{-N D_p(p_1,p_2,q_{\times},q_{\times}')}
}{N (p_2^2+(1-p_2)^2(1-p_1))+1}
$, where the exponential decreasing rate is given by
\begin{align*}
&D_p(p_1,p_2,q_{\times},q_{\times}') \\
:=&
(p_2^2+(1-p_2)^2(1-p_1))
h(
\frac{p_2^2 q_\times + (1-p_2)^2(1-p_1) q_{\times}'}{p_2^2 + (1-p_2)^2(1-p_1)})
\\
&-p_2^2 h(q_\times) - (1-p_2)^2(1-p_1) h(q_{\times}').
\end{align*}
When the phase error rate among raw keys is $q_{\times}'$,
the random privacy amplification with the sacrificed bit rate $S_2$
reduces
the block error probability of final keys in the $\times$ basis to 
$\delta_p(q_{\times}'):=
2^{-N [S_2-(1-p_2)^2(1-p_1)h(q_{\times}')]_+}$\cite{H}, where
$[x]_+$ is $x$ for a positive number $x$ while 
$[x]_+$ is zero for a negative number $x$.
Since $q_{\times}'$ takes the values in $\{0,1/N,2/N,\ldots, 1\}$, 
the (block) error probability of the final keys in the $\times$ basis is upperly bounded by
$B_p(N,p_1,p_2,q_{\times}):=
\sum_{k=0}^N \epsilon_p(\frac{k}{N}) \delta_p(\frac{k}{N})$.
Hence, applying the type method to the parameter $q_{\times}'$\cite{CK}, we obtain its exponential decreasing rate (See Hayashi\cite{H}):
\begin{align}
&\lim_{N\to \infty}\frac{-1}{N}\log B_p(N,p_1,p_2,q_{\times}) 
\nonumber \\
=&\min_{0\le q_{\times}' \le 1/2}
\Bigl( 
[S_2-(1-p_2)^2(1-p_1)h(q_{\times}')]_+
\nonumber \\
&+D_p(p_1,p_2,q_{\times},q_{\times}')
\Bigr)
\label{e4}.
\end{align}
In the following discussion, 
$S_2(p_1,p_2,q_{\times},C)$ presents the solution $S_2$ of $(\ref{e4})=C$.
In fact, 
Eve's Holevo information $\chi_{E}$
and Eve's distinguishability 
$\|\rho_{A,E}-\rho_{A}\otimes \rho_{E} \|_1$
are characterized by \cite{H,H2}
\begin{align*}
&{\rm E} \chi_{E}  \le (-\log B_p(N,p_1,p_2,q_{\times}) + M)\cdot
B_p(N,p_1,p_2,q_{\times})\\
&{\rm E} \|\rho_{A,E}-\rho_{A}\otimes \rho_{E} \|_1
 \le 
{\rm E} \max_{X}\|\rho_{E,X}- \rho_{E}\|_1 \\
& \le 
B_p(N,p_1,p_2,q_{\times}),
\end{align*}
where ${\rm E}$ denotes the average concerning random privacy amplification, 
$M$ denotes the length of the final keys, and
$\rho_{E,X}$ is Eve's state when the final key is $X$.
Hence, 
$B_p(N,p_1,p_2,q_{\times})$ can be regarded as an upper bound for 
Eve's distinguishability.
Thus, the quantity $S_2(p_1,p_2,q_{\times},C)$ is the minimum sacrificed bit rate 
in the random privacy under the condition that the exponential decreasing rate of the upper bound of Eve's distinguishability is greater than $C$.

Now, it will be shown why the rate $p_{2,A}$ of $\times$ basis of Alice is assumed to be equal to the rate $p_{2,B}$ of $\times$ basis of Bob.
If different rates are chosen, then
the performance is characterized by 
the coincidence probability 
in $+$basis $(1-p_{2,A})(1-p_{2,B})= 1+p_{2,A} p_{2,B}- 2(p_{2,A} + p_{2,B})$
and the coincidence probability in $\times$ basis $p_{2,A} p_{2,B}$.
Hence, it is sufficient to maximize the 
the coincidence probability 
in $+$basis $(1-p_{2,A})(1-p_{2,B})= 1+p_{2,A} p_{2,B}- 2(p_{2,A} + p_{2,B})$
with the condition that 
the coincidence probability 
in $\times$basis $p_{2,A} p_{2,B}$ is equal to an arbitrary constant $P$.
This maximum value is given when $p_{2,A}=p_{2,B}=\sqrt{P}$.
Thus, it is suitable to consider only the case of $p_{2,A}=p_{2,B}$.

In order to express the rate of sacrificed bits 
$S_2(p_1,p_2,q_{\times},C)$ as a function of the constraint $C$, we introduce two quantities, $q_{\times,1}'$ and $q_{\times,2}'$ as the solutions of the following equations in the range $[0,1/2]$: 
\begin{align*}
&D_p(p_1,p_2,q_{\times},q_{\times,1}')= C \\
&\frac{p_2^2 q_\times + (1-p_2)^2(1-p_1) q_{\times,2}'}
{p_2^2 (1-q_\times) + (1-p_2)^2(1-p_1) (1-q_{\times,2}')}
=(\frac{q_{\times,2}'}{1-q_{\times,2}'})^2.
\end{align*}
Then, the rate of sacrificed bits 
$S_2(p_1,p_2,q_{\times},C)$ is given as follows.
\begin{align*}
&S_2(p_1,p_2,q_{\times},C)
=
(1 - p_2)^2(1 - p_1)h(q_{\times,1}')
\end{align*}
when $q_{\times,1}' \le q_{\times,2}'$.
Otherwise,
\begin{align*}
S_2(p_1,p_2,q_{\times},C)
=
D_p(p_1,p_2,q_{\times},q_{\times,2}')+C.
\end{align*}

Next, we consider the (block) error probability of the final keys in the case when Gallager random coding and maximum likelihood decoding are applied\cite{Gal}.
When the bit error rate of the raw keys is $q_{+}'$
and the rate of sacrificed bits in classical error correction is $S_1$, the final error probability is upperly bounded by 
$\epsilon_b(q_{+}'):=
2^{-N [S_1-(1-p_2)^2(1-p_1)h(q_{+}')]_+}$.
We calculate the probability that the estimate of the bit error rate is $q_{+}$
and the phase error bit among raw keys is $q_{+}'$.
Similar to the case of the bit error rate,
by using the hypergeometric distribution,
this probability is upperly bounded by 
$\delta_b(q_{+}'):=\frac{
2^{-N D_b(p_1,p_2,q_{+},q_{+}')}
}{N(1-p_2)^2+1}
$,
where the exponential decreasing rate is given by
\begin{align*}
&D_b(p_1,p_2,q_{+},q_{+}') \\
:=&
(1-p_2)^2\Bigl(
h(p_1 q_+ + (1-p_1)q_+') 
-p_1h(q_+) \nonumber \\
&\hspace{10ex}- (1-p_1)h(q_+') \Bigr).
\end{align*}
Thus, the (block) error probability 
of the final keys in the $+$ basis is upperly bounded by
$B_b(N,p_1,p_2,q_+):=
\sum_{k=0}^{N} \epsilon_b(\frac{k}{N}) \delta_b(\frac{k}{N})$.
Applying the type method to the parameter $q_{+}'$\cite{CK}, we obtain the exponential decreasing rate:
\begin{align}
&\lim_{N\to \infty}\frac{-1}{N}\log B_b(N,p_1,p_2,q_+) \nonumber \\
=&\min_{0\le q_+' \le 1/2}
D_b(p_1,p_2,q_{+},q_{+}')
\nonumber \\
&
+[S_1-(1-p_2)^2(1-p_1)h(q_+')]_+\label{e3}.
\end{align}
In the following, 
$S_1(p_1,p_2,q_+,C)$ presents the solution $S_1$ of $(\ref{e3})=C$.
Thus, the quantity, $S_1(p_1,p_2,q_+,C)$, is the minimum sacrificed bit rate in the classical error correction under the condition that the exponential decreasing rate of the upper bound of error probability of the final keys is greater than $C$.

Similar to $S_2(p_1,p_2,q_{\times},C)$, in order to express the the rate of sacrificed bits  $S_1(p_1,p_2,q_{\times},C)$ as a function of the constraint $C$,
we introduce two quantities, $q_{+,1}'$ and $q_{+,2}'$ as the solutions of the following in the range $[0,1/2]$: 
\begin{align*}
& D_b(p_1,p_2,q_{+},q_{+,1}')=C \\
& \frac{p_1 q_+ + (1-p_1) q_{+,2}'}
{p_1 (1-q_+) + (1-p_1) (1-q_{+,2}')}
=(\frac{q_{+,2}'}{1-q_{+,2}'})^2.
\end{align*}
Therefore, 
$S_1(p_1,p_2,q_{\times},C)$ is given as a function of
the constraint $C$ as follows.
When $q_{+,1}' \le q_{+,2}'$,
\begin{align*}
S_1(p_1,p_2,q_{+},C)=
(1 - p_2)^2(1 - p_1)
h(q_{+,1}').
\end{align*}
Otherwise,
\begin{align*}
S_1(p_1,p_2,q_{+},C)
=
D_b(p_1,p_2,q_{+},q_{+,2}')+C.
\end{align*}
Hence, the final key generation rate
$R_{A}(p_1,p_2,q_+,q_{\times},C)$ is given by
\begin{align*}
&R_{A}(p_1,p_2,q_+,q_{\times},C)\\
=&
(1-p_2)^2(1-p_1)- S_1(p_1,p_2,q_{+},C)- S_2(p_1,p_2,q_{\times},C).
\end{align*}

Next, we consider the final key generation rate $R_{S}(p_1,q_+,q_{\times},C)$ in the symmetric case.
In this case,  
the exponential decreasing rate of the final error probability
is given by substituting 
$1/2$ into $p_2$ in the formula $S_1(p_1,p_2,q_+,C)$.
The exponential decreasing rate of Eve's distinguishability is given by substituting 
$1/2$ into $p_2$ in the formula $S_1(p_1,p_2,q_\times,C)$.
Thus, 
$R_{S}(p_1,q_+,q_{\times},C)$ is calculated by
\begin{align*}
&R_{S}(p_1,q_+,q_{\times},C)\\
=&\frac{1-p_1}{4}- S_1(p_1,1/2,q_+,E)- S_1(p_1,1/2,q_\times,C).
\end{align*}

Therefore, applying numerical analysis to these formulae, we obtain 
Figs. \ref{rate}, \ref{argmax1}, and \ref{argmax2}.

\vspace{2ex}
\noindent
[Discussion]\par
%\section{Discussion}
It has been shown that the asymmetric protocol improves the symmetric protocol under an exponential constraint condition based on the analysis on Hayashi\cite{H}.
This result suggests the importance of the choice of the ratio between the two bases when designing QKD system.
A similar result can be expected based on Lo et al \cite{Lo} and Scarani\cite{SR}.
It is interesting to compare the obtained result with those based on Lo et al \cite{Lo} and Scarani and Renner\cite{SR}.
A similar result can be expected in the decoy method\cite{hwang,wang1,LMC,H3}.
Future work will investigate the same problem in the finite-length framework using the decoy method \cite{HHHT}.
As well, it has been shown in this letter that the 
exponential rate is a useful criterion for the case of limited coding length.
It is interesting to apply this criterion to other topics in QKD.

\vspace{2ex}
\noindent
[Acknowledgement]\par
This research
was partially supported by the SCOPE project of the MIC of Japan and a Grant-in-Aid for Scientific Research on Priority Area `Deepening and Expansion of Statistical Mechanical Informatics (DEX-SMI)', no. 18079014.

\end{document}